\documentstyle[prd, preprint,eqsecnum,aps]{revtex}  
\begin{document}
\draft

\preprint{\vbox{
\hbox{CTP-TAMU-15/99, SINP-TNP/99-6}
\hbox{hep-th/9904129}
}}
  
\title{Non-Threshold (F, Dp) Bound States \\}

\author {J. X. Lu$^1$ and Shibaji Roy$^2$}
\address{$^1$Center for Theoretical Physics, 
 Texas A\&M University, College Station, TX 77843\\
E-mail: jxlu@rainbow.physics.tamu.edu\\
 $^2$Saha Institute of Nuclear Physics,
1/AF Bidhannagar, Calcutta 700 064, India\\E-mail: 
roy@tnp.saha.ernet.in}

\maketitle
\begin{abstract}

In the previous paper [hep-th/9904112], we argued that there exist 
BPS bound states of Dp branes
 carrying certain units of quantized constant electric field
 for every $p$ (with $1 \le p \le 8$). Each of these bound states
 preserves one half of the spacetime supersymmetries.
In this paper, we  construct these bound
state configurations explicitly for $2 \le p \le 7$ 
from Schwarz's $(m,n)$-string or (F, D1) bound state in type IIB 
string theory 
by T-dualities along the transverse directions. We calculate 
the charge per 
$(2\pi)^{p -1}   \alpha'^{(p - 1)/2}$  of $(p -1)$-dimensional area 
for F-strings in 
(F, Dp) and the tension for each of these bound states. 
The  results agree
precisely  with those obtained previously from the 
worldvolume study. We  study the 
decoupling limit for the (F, D3) bound state and 
find that Maldacena's 
$AdS_5/CFT_4$ correspondence may hold true even with 
respect to this
bound state but now with an effective string coupling 
rather than the usual string coupling.
This coupling  is quantized and can be independent of 
the usual string coupling in certain limit.
\end{abstract}
\newpage

\section{Introduction\protect\\}
\label{sec:intro}
In the absence of a complete formulation of M/string
theory, BPS $p$-brane solutions from various supergravities, 
which are the low 
energy and/or weak-coupling limits of this intrinsically 
non-perturbative unified
theory, are almost the only sources from which reliable non-perturbative 
information about this theory can be extracted. These solutions played
vital role in every major development in the past and will continue to do so.
BPS properties obtained in the low energy and weak-coupling limits 
remain to be valid
non-perturbatively for the corresponding BPS states. 
These properties should also be
independent of whether they are obtained from fields in bulk 
spacetime or from fields on
the corresponding worldvolume. For this reason, BPS properties 
obtained from the 
supergravity $p$-brane 
solutions are often used to obtain information about fields living on the 
corresponding 
worldvolume and vice-versa. For example, the mass per unit $p$-brane 
volume calculated from a spacetime  BPS Dp-brane solution should 
correspond to the Dp-brane 
tension of the corresponding Born-Infeld action. This is trivial for simple 
Dp-brane
solutions. But it is quite non-trivial for complicated solutions such as BPS
bound state solutions. For example, the tension formula of 
$(m, n)$-string bound state of Schwarz \cite{schone}, with $m, n$ 
relatively prime integers,
 can be used to determine the corresponding D-string
worldsheet $U(1)$ field strength, in the linear approximation, 
as $g m$ with $g$ the string coupling
\cite{calm}. This in turn can be used to determine the point charge due  to 
the ending of a fundamental string (for short, F-string) on a general Dp-brane.

	After all, a $p$-brane configuration represents the field configuration
created by the corresponding $p$-brane source just like a static electric field
due to a point charge represents the field surrounding the point source. Static 
interactions should be the same whether we calculate from the fields or from
the sources. As mentioned above, BPS
properties should be the same no matter whether they are obtained from the
field configuration or from the source. However, apart from these 
BPS properties, one must be careful in using them since in general 
the fields and 
the source are not independent and they actually interact with
each other through the so-called backreaction except in the case where the 
so-called decoupling limit is taken. In this limit, the modes propagating on 
the worldvolume will decouple from the modes propagating in the bulk spacetime 
even though these spacetime modes may have their origins from the worldvolume.
So we expect that in the decoupling limit either the worldvolume fields or
the bulk 
fields can be used to describe the same physics. Recently proposed $AdS/CFT$
correspondence by Maldacena \cite{mal} is one such example.

	So BPS $p$-brane solutions can provide not only the non-perturbative 
information but also a basis for new types of strong-weak dualities. These
are certainly important for us to seek the eventual formulation of M/string
theory, along with the powerful D-brane picture of Polchinski\cite{pol}.

	In this paper, we will construct new spacetime solutions for 
BPS bound states of Dp-branes carrying certain units of the 
quantized worldvolume 
constant electric field strength. 
We argued in our previous 
paper \cite{lurone} that such BPS bound states should exist for 
$ 1 \le p \le 8$ in both type IIA (when $p$ is even) and type IIB 
(when $p $ is odd) theories\footnote{
In this paper, we specialize in cases of $2 \le p \le 7$.}. The existence of 
these bound states was also discussed in \cite{arfs} but in a different approach
of mixed boundary conditions.
The non-vanishing $U(1)$ field 
strength on the worldvolume indicates that such a bound state 
carries information 
about F-strings ending on the Dp-brane, yet preserving one half 
of the spacetime 
supersymmetries. These Dp-brane bound states should be identified 
with the so-called
(F, Dp) bound states which can be obtained from the $(m,n)$-string of
Schwarz or (F, D1) bound state
in type IIB string theory by T-dualities along the transverse directions. 
Here integers 
$m, n$ are relatively prime. This is precisely the method which 
we will use here to construct
these solutions which was also mentioned in \cite{rust,cosp}.
 The $p = 3, 4, 6$ configurations for (F, Dp) have been given
in \cite{rust,grelpt,cosp}, respectively. Other non-threshold bound states for
one $p'$-brane within another p-brane with $p' < p$ in M and Type II
string theories have been discussed in \cite{rust,grelpt,cosp,col}. 

 Our worldvolume picture of these Dp-brane bound states 
indicates clearly that the notation `F' in (F, Dp) represents actually 
an infinite number of parallel  NS-strings. 
Precisely, we have
one NS-string per $(2\pi)^{p - 1} \alpha'^{(p - 1)/2}$ area 
over a $(p - 1)$-dimensional plane which is perpendicular to  
this string. 
Each NS-string is  $m$ F-strings if the quantized worldvolume 
constant field strength
$F_{01} = g m$, with $g$ the string coupling constant (where the field 
lines are chosen along 
$x^1$ axis). These were obtained in \cite{lurone} by noting that 
the (F, D1) worldvolume action can be obtained from a (F, Dp), for 
$ p \ge 2$, worldvolume
action by T-dualities.  By this, we also determined the tensions for 
(F, Dp) bound states
 in \cite{lurone}
which clearly indicates that (F, Dp) is a non-threshold bound state. 
We will show in 
this paper from the corresponding 
spacetime solutions that indeed these are all true.

Since we have constructed in this paper the space-time configurations
for all the (F, Dp) bound states in type IIA and IIB theories, it would
be natural to look at how they affect the $AdS/CFT$ correspondences
conjectured by Maldacena. By examining the (F, D3) configuration carefully,
we find that the decoupling limit in this case does not automatically
imply the space-time geometry to be $AdS_5 \times S^5$ as happens in the
case of Maldacena, i.e., for simple D3-branes. We also find that the string
coupling constant in this case is replaced by a finite but a smaller 
effective string coupling constant. However, we observed that by a suitable
rescaling of the coordinates the $AdS_5/CFT_4$ correspondence may still hold
true, but now with respect to (F, D3) bound state rather than the simple
D3-branes. The effective string coupling constant in the near horizon region
is quantized in terms of the integer $n$, the 5-form flux and the integer
$m$, the number of F strings per $(2\pi)^2 \alpha'$ area over the two
dimensional plane perpendicular to the F strings, where $m$, $n$ are
relatively prime integers. This coupling can be independent of the usual
string coupling at spatial infinity in the limit $m \gg n$.

	This paper is organized as follows. In the following section, 
we discuss in 
some detail
the so-called vertical and diagonal (or double) dimensional 
reductions/oxidations
 in type 
IIA and type IIB supergravities which are the basic methods we use 
to construct 
the (F, Dp) configurations. In section 3, we  first give a detail 
construction of 
(F, D2) as an illustrating example. We then list results for each of 
(F, Dp) for 
$ 3 \le p \le 7$. We also present the (W, Dp) (with W representing the waves)
solutions for $0 \le p \le 6$
 from these newly found (F, Dp) configurations by T-dualities along 
the direction
 of F-strings. In section 4, we calculate the charge per unit
$(p - 1)$-dimensional area for the F-strings and show that we indeed have 
$m$ F-strings 
per $(2\pi)^{p -1} \alpha'^{(p - 1)/2}$ of ($p - 1$)-dimensional area for 
these bound states.
We also calculate the mass per unit $p$-brane volume for these 
solutions and show that the 
corresponding tensions agree precisely with what we obtained in 
\cite{lurone} based on
the worldvolume study. In section 5, we study the decoupling limit for the
(F, D3) bound state. Finally, we conclude this paper in section 6.

\section{Diagonal and Vertical Dimensional Reductions/Oxidations\protect\\}
\label{sec:dvdr}
It is well-known that type IIA and type  IIB superstring theories 
compactified on a circle are equivalent by T-duality i.e.
order by order in perturbation theory. More precisely, 
IIA and IIB theories are interchanged by  the T-duality transformation 
$R \rightarrow 1/R$, with $R$ the compactified radius, along with 
an interchange of 
momentum or KK modes with the winding modes. However, this equivalence 
does not extend to the 
respective $S^1$-compactified supergravity theories due  to the absence of the 
string winding modes.  But this will not have any effect on the massless 
modes in $D = 9$.
Therefore, the $D = 9, N = 2$ supergravity obtained from type IIA supergravity 
by the 
dimensional reduction must be equivalent by T-duality to that obtained 
from type IIB supergravity. This  also
follows from supersymmetry because $D = 9, N = 2$ supergravity is unique 
up to field
redefinitions. Hence, the T-duality acting on the $D = 9$ massless 
fields must map
fields in IIA basis to those in IIB basis. This feature, as we will see, 
is  precisely what we 
need to obtain a BPS solution in type IIA theory from a BPS solution in 
type IIB theory and vice versa by 
a T-duality transformation.
     
It is well-known that type II theories contain the following simple
BPS states\footnote{
By simple, we mean an object whose charge is associated with only 
one antisymmetric 
tensor field in 
type II theories. We have not included here the D9 branes, i.e., the 
so-called spacetime-filling branes\cite{ber}.}:
\begin{eqnarray}
{\rm Type~  IIA}~\qquad&& {\rm W}~~{\rm F}~~{\rm NS5}~~{\rm KK}~~
\qquad~{\rm D0}~~{\rm D2}~~{\rm D4}~~{\rm D6}~~{\rm D8}\nonumber\\
{\rm Type~  IIB}~\qquad&& {\rm W}~~{\rm F}~~{\rm NS5}~~{\rm KK}~
\qquad~{\rm D(-1)}~~{\rm D1}~~{\rm D3}~~{\rm D5}~~{\rm D7}\nonumber\\
\end{eqnarray}
where W, F, NS5 and KK  denote waves, fundamental strings,  NS fivebranes, 
and KK
monopoles, respectively, and they  are associated with the NSNS fields. 
Also, Dp ($ - 1\le p \le 8$)
are the so-called D-branes and they are associated with  the so-called 
RR fields.
 The action of T-duality along parallel or transverse direction 
 on the above objects present in type II theories
is given in the following table. 

\begin{center}
 \begin{tabular}{|c|c|c|}  \hline
  & Parallel
 & Transverse\\
 \hline
Dp& D(p $-$ 1) & D(p + 1)\\
F& W  & F \\
W& F  & W \\
NS5& NS5 &KK \\
KK & KK & NS5 \\
\hline\end{tabular}
\end{center}

The T-duality may be performed along either one of the 
worldvolume directions or one of the transverse directions 
(for KK monopole, the transverse 
direction is taken to be the nut direction).

It is clear from the above table that if we start from a (F, D1) bound 
state in
the type IIB theory, we will have (F, D2) bound state in type IIA theory by
T-dualizing (F, D1) along one of its transverse directions. Then we can have 
(F, D3) bound state in the type IIB theory by T-dualizing (F, D2) along one of
its transverse directions. Repeating this procedure, we can have (F, D4) from
(F, D3), (F, D5) from (F, D4), (F, D6) from (F, D5) and finally (F, D7) from
(F, D6). We cannot obtain (F, D8) from (F, D7) by this procedure. We will
explain possible reasons behind this at the end of this section. 
If we T-dualize 
the above bound
states along the longitudinal direction of F in (F, Dp) for $ 1 \le p \le 7$, 
we have (W, Dp) bound states for $ 0 \le p \le 6$ with W 
representing the waves.

The question now is how to implement the above T-duality 
at the level of solutions of the IIA and IIB supergravities. 
As discussed above,
the $S^1$ compactified IIA and IIB supergravities are not equivalent by
T-duality except for the massless sector of $D = 9, N = 2$ supergravity. 
This implies that 
T-duality can be implemented in obtaining solutions in one supergravity
from the known solutions each with one isometry along the 
would-be compactified direction in the other supergravity. This is exactly the 
rationale behind the procedure of the so-called vertical dimensional reduction
and diagonal (or double) dimensional oxidation or the diagonal reduction and 
vertical oxidation, depending on whether the T-duality acts along a transverse
direction or a worldvolume direction. 

In this paper, we choose a convention such that $H_n$ denotes an
 $n$-form field
strength in type IIB supergravity or in $D = 9, N = 2$ supergravity 
in terms of the  
type IIB basis
and similarly $F_n$ denotes an $n$-form in type IIA supergravity or in 
$D = 9, N = 2$ 
supergravity 
in terms of the type IIA basis. If an $n$-form is reduced to 
an $(n-1)$-form 
from $D = 10$ to $D = 9$, we denote the resulting $(n-1)$-from with 
a superscript `1'
indicating that it is obtained by one-step reduction. 
For example, if the reduction is
along one of the indices of $H_n$,  we have 
$H_n \rightarrow H^{(1)}_{n - 1} \wedge dz$.

In the case of T-duality along a transverse
direction of a BPS $p$-brane solution ($ p \le 6$) as discussed in 
\cite{dabghw,lups}, we 
first 
use the ``no-force" condition
for the $p$-brane configuration to obtain a multi-center solution from a 
single center 
one by placing many $p$-branes parallel to each other. Without loss 
of generality and to be 
specific, let us assume that this $p$-brane is in type IIB supergravity. 
The multi-center solution can be obtained by
the following replacement in the corresponding 
Harmonic function. For example, if a single-center Harmonic function 
$H = 1 + Q_p /r^{7 - p}$ with $r^2 = (y^1)^2 + (y^2)^2 + \cdots + 
(y^{9 - p})^2$, then the multi-center Harmonic function will be given as, 
\begin{equation}
H = 1 + \sum_n \frac{Q_p^{(n)}}{\mid \vec{y} - \vec{y}_n \mid^{7 - p}}.
\label{eq:mcs}
\end{equation}
In order to generate one isometry along the T-dual direction, say, 
$y^{9 - p}$-direction, as pointed out in \cite{lups}, a continuous 
stack of $p$-branes with uniform charge density should be placed along the 
coordinate $z = y^{9 - p}$ such that the summation in 
Eq.\ (\ref{eq:mcs}) can be replaced by an integration\footnote{This seems
to work also for non-BPS solutions as discussed in \cite{lups}.}. For 
the purpose
of this paper, let us take a close look at this continuum limit. We first
 take $\vec{y}_n = 2 \pi n a \hat{z}$ in Eq.\ (\ref{eq:mcs}), 
with $n \in Z$,
$\hat{z}$ the unit
vector along $y^{9 - p}$-direction and $a =  \sqrt{\alpha'}$ where $\alpha'$
is the string constant.  If we allow the
range of the summation in Eq.\ (\ref{eq:mcs}) to  be 
$-\infty < n < \infty$ and 
take all charges to be equal, i.e., $Q_p^{(n)} = Q_p$, then the 
multi-center solution
is an infinite periodic array of $p$-branes. The solution would be a 
stringy one if we take
$\alpha' \rightarrow 0$. Under this limit, the summation 
in Eq.\ (\ref{eq:mcs}) can also
be replaced by an integration, i.e.,
\begin{equation}
Q_p \sum_{n = - \infty}^\infty  \frac{1}{\mid \vec{y} - 2\pi n a 
\hat{z} \mid^{7 - p}} 
\rightarrow \frac{Q_p}{2\pi a} \int_{- \infty}^\infty \frac{dz}
{(r^2 + z^2)^{(7 - p)/2}}
= \frac{\tilde{Q}_p}{ r^{6 - p}},
\end{equation}
where $\tilde{Q}_p = c_p Q_p$ with constant $c_p = Q_{p + 1}/Q_p$.  
The explicit
expressions for $Q_{p + 1}$ and $Q_p$ are given in the next section. 
In the above,
we have now  $r^2 = (y^1)^2 + \cdots + (y^{8 - p})^2$. In other
words, one isometry is generated along the T-dual direction by the 
infinite periodic array
of $p$-branes. Once this solution is known, a BPS $p$-brane solution in 
$D = 9$ can be obtained from 
it. The Einstein metric for $D = 9$ $p$-brane can be obtained from 
the corresponding metric of
$D = 10$ $p$-brane as (see, for example, the appendix of  \cite{duflp})
\begin{equation}
ds^2_{10}~ ({\rm type~ IIB})= e^{-\varphi_B/(2\sqrt{7})} ds_9^2 + 
e^{\sqrt{7} \varphi_B/2} d z^2,
\label{eq:mriib}
\end{equation}
where $\varphi_B$ is a dilatonic scalar originating from the 
dimensional reduction.
Since we know $d s^2_{10}~({\rm type~ IIB})$ and it also has 
one isometry along 
$z$-direction, 
we can read from 
the above
the new field $\varphi_B$ and therefore the $D = 9$ Einstein metric 
$d s_9^2$. The $D = 10$ dilaton 
$\phi_B$ remains the
same as the one for the solution of the infinite periodic array of 
$p$-branes in $D = 10$. The rule for reducing 
an $n$-form
field strength $H_n$ is as follows: if it carries an electric-like charge, 
it will remain the same 
in $D = 9$
whereas it will reduce to an $(n-1)$-form, i.e., 
$H_n \rightarrow H^{(1)}_{n - 1}\wedge dz$,
 if it carries a magnetic charge. So we now obtain a $D = 9$ BPS 
$p$-brane solution from a $D = 10$ BPS $p$-brane solution by the 
so-called vertical dimensional reduction.

We expect that the T-dual of the above $p$-brane in type IIB gives a 
$(p + 1)$-brane in type IIA. 
Let us first pretend that we know this $(p + 1)$-brane solution in type IIA. 
We denote the brane coordinate
$x^{ p + 1} = z$ ( We choose the brane along $x^1, \cdots, 
x^{p + 1}$-directions). All the fields for a
static BPS $(p + 1)$-brane are usually independent of the brane coordinates. 
So we have one isometry already
along the $z$-direction. Thus a BPS $p$-brane solution in $D = 9$ 
can be obtained simply by
 the so-called diagonal or double dimensional reduction \cite{dufhis} on the 
BPS $(p + 1)$-brane solution of type IIA supergravity. 
The two Einstein metrics are now simply related to
each other by
\begin{equation}
d s^2_{10}~ ({\rm type~ IIA}) = e^{-\varphi_A/(2\sqrt{7})}ds_9^2 
+ e^{\sqrt{7} \varphi_A/2} d z^2,   
\label{eq:mriia}
\end{equation}
where $\varphi_A$ is another dilatonic scalar due to this 
dimensional reduction.
The IIA dilaton $\phi_A$ remains unchanged. The $n$-form field strength 
$F_n$ is reduced according to 
the following:
if it carries an electric charge, it will become an $(n-1)$-form in 
$D = 9$, i.e., 
$F_n \rightarrow F^{(1)}_{n -1} \wedge dz$, whereas it will remain the 
same if
it carries a magnetic charge, just the opposite to the case of 
vertical reduction. Here we do not really
change anything. It is just the same whether we call the solution a 
$(p + 1)$-brane in $D = 10$ or a $p$-brane
in $D = 9$. In other words, if we know a $p$-brane solution in $D = 9$, 
we should obtain a $(p + 1)$-brane solution in
$D = 10$ right away and vice-versa. The process of obtaining a 
$(p + 1)$-brane in $D = 10$ from a $p$-brane 
in $D = 9$ is called the diagonal oxidation. 

The $p$-brane solution in $D = 9$ obtained from a known type IIB $p$-brane 
solution is described  in terms of
fields in type IIB basis while in obtaining a $(p + 1)$-brane in type IIA, 
we need a $p$-brane solution in $D = 9$
described  in terms of the fields in type IIA basis. So we need to map 
the fields in type IIB basis to fields
in type IIA basis. This mapping is nothing but the T-duality transformation 
for the solution. This is also
the T-duality transformation acting on the fields of $D = 9, N = 2$ 
supergravity 
discussed earlier. Then T-duality on field strengths in $D = 9, N = 2$
supergravity or the corresponding $p$-brane solutions is the identification 
$H_n  = F_n$ with the understanding that if $H_n$ is defined with 
a Chern-Simons term
so is $F_n$. The mapping for the $D = 10$ dilaton $\phi$ and the dilatonic 
scalar $\varphi$ between the two versions of $D = 9, N = 2$ supergravity is
\begin{equation}
\left(\begin{array}{c}
       \phi_A\\
       \varphi_A \end{array}\right) 
         = \left(\begin{array}{cc}
           \frac{3}{4}& - \frac{\sqrt{7}}{4}\\
           - \frac{\sqrt{7}}{4} & - \frac{3}{4} \end{array}\right)
          \left(\begin{array}{c}
       \phi_B\\
       \varphi_B \end{array}\right).
\label{eq:tdvdr}
\end{equation}
                                    
The above relations have been given or can be deduced, for example, from
the appendix in \cite{duflp} where the relations between the corresponding 
gauge
potentials are also given\footnote{The precise relations between various fields in two different 
bases of the nine dimensional theory are given in a recent 
paper \cite{lupsone}. We would like to thank Chris Pope for letting
us know their results prior to publication. These relations are also given in
\cite{berho} but in a different conventions than we use here.}.

Using the above relations, we can obtain a $p$-brane solution in $D = 9$ 
in type IIA basis from a $p$-brane solution in type IIB basis. 
Then we obtain 
a BPS $(p + 1)$-brane solution in type IIA by oxidizing 
this $p$-brane solution back to $D = 10$ as described above.

If we reverse the above process, we then obtain a BPS $p$-brane solution 
in one
theory from a known BPS $(p + 1)$-brane solution in the other theory now by
a T-duality along a world volume direction. The oxidation involved in this 
process is the so-called vertical one\cite{lups}.

	Now we conclude this section by explaining why we cannot 
obtain D8 brane
solution in the massive type IIA supergravity from a D7 solution in the type 
IIB supergravity by a T-duality transformation along one of its  transverse
directions. First, it is not clear even now whether the massive type IIA
supergravity has its origin in any of the known superstring theories or
even in M-theory (though we believe so). So it is not clear that there exists 
a theory from which the massive type IIA supergravity originates and is T-dual
to the type IIB superstring theory. Even if we have such  a theory, 
the massless
fields in $D = 9$ from the $S^1$-compactified massive type IIA supergravity
cannot describe the 8-brane in $D = 9$. Therefore, we do not expect that the
D8-brane solution is related to a D7 brane solution in the type IIB 
supergravity
by a T-duality transformation\footnote{We merely mean this at the level of 
supergravity solutions. As we know, in string theory, D8 brane does seem to
be related to D7 brane by 
a T-duality.}. Technically, as explained in \cite{lups}, the
vertical reduction does not seem to apply when $p = 7$. By the same token, we
cannot obtain (F, D8) solution from (F, D7) solution by a T-duality 
transformation.

\section{(F, D$_p$) Bound States\protect\\}
\label{sec:fdpbs}

	In this section, we will construct explicitly the (F, Dp) bound state
configurations for $ 2 \le p \le 7$ from the known (F, D1) bound state
configuration of Schwarz \cite{schone} by a T-duality transformation 
along one of the transverse directions described in the previous section. 
We will give a detail construction for the case of (F, D2) as an example and
present only the results  for the rest. Once we have (F, Dp) configurations for
$ 1\le p \le 7$, we will also present the results for (W, Dp) for 
$0 \le p \le 6$, i.e., a Dp brane
carrying waves in it, by a T-duality along the longitudinal direction 
of F-strings.
All these configurations preserve one half of the spacetime supersymmetries.

	The field configurations of the $(m,n)$-string of Schwarz 
(or (F, D1) bound state), with
$m, n$ relatively prime integers, in type IIB theory \cite{schone}
are given in terms of our notations as: the Einstein metric,
\begin{equation}
ds^2~ ({\rm type ~IIB}) = H^{-3/4} [ - (d x^0)^2 + (d x^1)^2 ] 
+ H^{1/4} dy^i dy^i,
\label{eq:fd1m}
\end{equation}
with $i = 1, \cdots, 8$; the type IIB dilaton,
\begin{equation}
e^{\phi_B} = e^{\phi_{B0}}\, H' H^{-1/2},
\label{eq:iibd}
\end{equation}
the axion,
\begin{equation}
\chi_B = \frac{mn (H - 1) + \chi_{B0} \,\Delta_{(m, n)} \,e^{\phi_{B0}}}
              { n^2 H +  (m - \chi_{B0} n)^2 \,e^{2\phi_{B0}}},	  
\label{eq:iiba}
\end{equation}
 and the NSNS 3-form field strength $H_3 ~({\rm NSNS})$ and 
RR 3-form $H_3~({\rm RR})$,
\begin{eqnarray}
H_3~ ({\rm NSNS}) &=& - \Delta_{(m, n)}^{-1/2} \,e^{\phi_{B0}} 
\,(m - \chi_{B0} n) d H^{-1}
\wedge d x^0 \wedge d x^1,\nonumber\\
H_3~ ({\rm RR})& = &\Delta_{(m, n)}^{-1/2} \left[  \chi_{B0}\, 
(m - \chi_{B0} n)\,
e^{\phi_{B0}} -  n \,e^{-\phi_{B0}}\right] d H^{-1}\wedge d x^0 \wedge d x^1.
\label{eq:3f}
\end{eqnarray}

In the above, $\phi_{B0}$ and $\chi_{B0}$ represent the asymptotic 
values of the 
type IIB dilaton and axion, respectively.  The SL(2, Z) invariant 
$\Delta$-factor is
\begin{equation}
\Delta_{(m, n)} = (m - \chi_{B0} n)^2 e^{\phi_{B0}} +  n^2 e^{-\phi_{B0}},
\label{eq:df}
\end{equation}
and the SL(2, Z) invariant Harmonic function $H$ is 
\begin{equation}
H = 1 + \frac{Q_1}{r^6},
\label{eq:hf1}
\end{equation}
where the radial distance $r^2 = (y^1)^2 + (y^2)^2 + \cdots + (y^8)^2$ and
the quantized central charge $Q_1$ is
\begin{equation}
Q_1 =   2^5 \pi^2 \alpha'^3 \Delta_{(m, n)}^{1/2},
\label{eq:cc1}
\end{equation}
with $\alpha'$ the string constant. We have also introduced a second 
Harmonic function
\begin{equation}
H' = \frac{ (m - \chi_{B0} n)^2 e^{\phi_{B0}} +  n^2 H e^{-\phi_{B0}}} 
{\Delta_{(m, n)}},
\label{eq:nhf2}
\end{equation}
which approaches unity as $r \rightarrow \infty$.

	Before we move on to the constructions of (F, Dp) bound states, 
we fix a few
conventions. The RR-charge of the Dp-brane in (F, Dp) is defined 
\cite{dufkl} as,
\begin{equation}
e_p = \frac{1}{\sqrt{2} \kappa_0} \int_{S^{8 - p}} e^{ - a(p) 
\phi} \ast G_{p
+2},
\label{eq:ecd}
\end{equation}
for Noether ``electric" charge, and
\begin{equation}
 g_p = \frac{1}{\sqrt{2} \kappa_0} \int_{S^{p + 2}} G_{p + 2},
\label{eq:mcd}
\end{equation}
for topological or magnetic-like  charge. In the above $\sqrt{2} \kappa_0 =
(2\pi)^{7/2} \alpha'^2$. For a Noether charge,
the integration is over an asymptotic $(8 - p)$-sphere surrounding the 
Dp-brane
while for a magnetic-like charge, the integration is over an asymptotic 
$(p + 2)$-sphere surrounding the Dp-brane. Also, $G_{p + 2} = H_{p + 2}$ 
for type IIB
theory while $G_{p + 2} = F_{p + 2}$ for type IIA theory. 
 The constant $a (p) = (p - 3)/2$ for Dp-branes in  both type IIA 
and type IIB theories. The above definitions are also valid for NSNS branes,
i.e., NSNS strings and fivebranes, but with $a (p) = - (p - 3)/2$. 
However, the
charge associated with the F-strings in the (F, Dp) bound states cannot be
calculated using the above simple formula. We will discuss this in the next
section. In this paper, $\ast$  always  denotes the Hodge dual. $\epsilon_n$
denotes the volume form on an $n$-sphere where the volume of 
a unit $n$-sphere is
\begin{equation}
\Omega_n = \frac{2 \pi^{(n + 1)/2}}{\Gamma ((n + 1)/2)},
\label{eq:usv}
\end{equation}
and the unit charge
for a Dp-brane is given as,
\begin{equation}
Q_0^p \equiv  (2\pi)^{(7 - 2p)/2} \alpha'^{(3 - p)/2}.
\label{eq:puc}
\end{equation}

{\bf (F, D2) Bound State}: Now in order to construct this bound state we 
first need to construct, as described in the previous section, the solution
corresponding to an infinite periodic array of $(m,n)$-strings along 
$z = y^8$ axis. 
The Harmonic function in that case would be given as,
\begin{equation}
H = 1 + Q_1 \sum_{n = -\infty}^\infty \frac{1}{\mid \vec{y} 
- 2\pi n a \hat{z}\mid^6},
\label{eq:mcs1}
\end{equation}
where $\hat{z}$ denotes the unit vector and $a = \sqrt{\alpha'}$. 
In the limit $\alpha' \rightarrow 0$, the
summation in the above Harmonic function can be replaced by an 
integration, i.e.,
\begin{equation}
\sum_{n = -\infty}^\infty \rightarrow \int_{-\infty}^\infty \frac{dz}{2\pi a}.
\label{eq:sir}
\end{equation} 

Note that writing $\vec{y} = \tilde{\vec{y}} + z \hat{z}$ and so,
$\mid \vec{y} - 2\pi na \hat{z} \mid^6 = [ {\tilde{r}}^2 
+ (z - 2\pi na)^2]^3$, with
${\tilde r}^2 = (y^1)^2 + \cdots + (y^7)^2$, we have,
\begin{equation}
\sum_{n = -\infty}^\infty \frac{1}{\mid \vec{y} - 2 
\pi n a \hat{z}\mid^6}
\rightarrow \int_{-\infty}^\infty \frac{dz}{2\pi a} \frac{1}
{({\tilde r}^2 +
z^2)^3} = \frac{3}{16 a} \frac{1}{{\tilde r}^5}.
\label{eq:ir1}
\end{equation}
  
 So the new Harmonic function describing the infinite array of $(m,n)$-strings
 along the $z$-direction is 
 \begin{equation}
 H = 1 + \frac{Q_2}{r^5},
 \label{eq:nh2}
 \end{equation}
 where we have dropped the `tilde' above the radial distance and 
will continue to do so
from now on. The
 central charge $Q_2$ in (3.16) is
 \begin{equation}
 Q_2 = \frac{3 Q_1}{16 \alpha'^{1/2}} = 6 \pi^2 \alpha'^{5/2} 
 \Delta_{(m, n)}^{1/2},
 \label{eq:cc2}
 \end{equation}
The Harmonic function is now independent of the $z$-coordinate. 
In other words,
the Einstein metric (Eq.\ (\ref{eq:fd1m})) expressed in terms of this 
Harmonic 
function possesses one additional isometry along the $z$-direction. If we now
compactify the metric along $z$-direction, Eq.\ (\ref{eq:mriib}) will give
\begin{equation}
e^{\sqrt{7} \varphi_B /2} = H^{1/4},
\label{eq:vd2}
\end{equation}
and the 9-d Einstein metric will be given as,
\begin{equation}
d s^2_9 = H^{- 5/7} [ - (d x^0)^2 + (d x^1)^2 ] + H^{2/7} dy^i dy^i,
\label{eq:m9}
\end{equation}
where  $i = 1, 2, \cdots, 7$. For the 9-dimensional solution, the type IIB
dilaton, axion and NSNS and RR 3-forms remain unchanged with the 
replacement of the
old Harmonic function by the new one. We have therefore obtained 
$(m,n)$-string bound
state in $D = 9$ in the type IIB basis. 

We now express all the relevant fields in type IIA basis. The $d s_9^2$ 
remains 
unchanged. $F_3 = H_3 ~({\rm NSNS})$, $F_3^{(1)} = H_3 ~({\rm RR})$ and 
$F_1^{(1)} = d \chi$. The type IIA dilaton $\phi_A$ and the dilatonic scalar
$\varphi_A$ can be obtained from $\phi_B$ and $\varphi_B$ through 
Eq.\ (\ref{eq:tdvdr}) as
\begin{eqnarray}
e^{\phi_A} &=&  e^{3\phi_{B0}/4} H'^{3/4} H^{- 1/2},\nonumber\\
e^{\sqrt{7} \varphi_A} &=& e^{-7 \phi_{B0}/4} H'^{- 7/4} H^{1/2},
\label{eq:iiadvd}
\end{eqnarray}
where the Harmonic function $H'$ continues to be given by the expression 
\ (\ref{eq:nhf2}) but in terms of  the present Harmonic function  $H$.

	Once we express  $D = 9$ (F, D1) configuration in the type IIA
basis, we can read the (F, D2) configuration in type IIA theory using the 
diagonal oxidation described in the previous section. For example, the 
Einstein metric in $D = 10$ for this configuration can be read from 
Eq.\ (\ref{eq:mriia}) with $d s^2_9$, $\phi_A$ and
$\varphi_A$ given in (3.19) and (3.20). We here collect the complete 
results for the (F, D2)
configuration for the metric,
\begin{equation}
d s^2_{10}~ ({\rm type~ IIA}) = e^{\phi_{B0}/8} 
H'^{1/8} H^{1/4} \left[ H^{-1} 
(- (d x^0)^2 + (d x^1)^2 ) + e^{ - \phi_{B0}} H'^{ - 1} (dx^2)^2
+ d y^i d y^i\right],
\label{eq:iiadtwom}
\end{equation}
for the dilaton,
\begin{equation}
e^{\phi_A} = e^{3\phi_{B0}/4} H'^{3/4} H^{-1/2},
\label{eq:iiadtwod}
\end{equation}
 and for the remaining non-vanishing fields\footnote{In type IIA theory, 
$F_4' = dA_3 + A_1\wedge F_3$},
\begin{eqnarray}
F_2 &=& n \Delta_{(m,n)}^{-1} ~e^{-\phi_{B0}} 
(m -\chi_{B0} n)~ H'^{-2} ~d H\wedge
dx^2,\nonumber\\
F_3 &=& - \Delta_{(m,n)}^{-1/2} ~e^{\phi_{B0}} ~(m - \chi_{B0} n) 
~d H^{-1}\wedge
d x^0\wedge d x^1,\nonumber\\
F_4' &=& n \,e^{- 3 \phi_{B0}/8}\, H'^{-3/8} \, H^{1/4}\, 
\frac{\sqrt{2} \kappa_0 Q_0^2}
{\Omega_5} \,\ast \epsilon_6.
\label{eq:tstwo}
\end{eqnarray}
 In the above,  $i =  1, 2, \cdots
7$, the Harmonic functions $H'$ and $H$ are given by 
Eqs.\ (\ref{eq:nhf2}) and (\ref{eq:nh2}), respectively.

{\bf (F, D3) Bound State}\footnote{This configuration with zero 
asymptotic values of
$\phi_{B0}$ and $\chi_{B0}$ was also given in
\cite{rust} for different purpose. Some non-threshold bound 
states of M theory, 
not considered here, were also discussed there.}: Once 
we obtain (F, D2), we can repeat the above
process to obtain (F, D3) bound state configuration. The results are  for
the Einstein metric,
\begin{eqnarray}
ds^2_{10}~ ({\rm type ~IIB}) = &&e^{\phi_{B0}/4} \,H'^{1/4}\, 
H^{1/4} [ H^{-1} (
- (d x^0)^2  + (d x^1)^2) \nonumber\\
&& + e^{-\phi_{B0}}\, H'^{-1} ( (d x^2)^2 + (d x^3)^2)
+ d y^i d y^i],
\label{eq:miibthree}
\end{eqnarray}
with $i = 1, 2, \cdots, 6$; for the type IIB dilaton,
\begin{equation}
e^{\phi_B} = e^{\phi_{B0}/2}\, H'^{1/2}\, H^{-1/2},
\label{eq:iibdthree}
\end{equation}
and for the remaining non-vanishing fields,
\begin{eqnarray}
H_3 ~({\rm NSNS}) &=& - \Delta_{(m,n)}^{-1/2} \,e^{\phi_{B0}} 
\,(m - \chi_{B0} n) \,
                    d H^{-1}\wedge d x^0 \wedge d x^1,\nonumber\\
H_3 ~({\rm RR}) &=& n \,\Delta_{(m,n)}^{-1}\, e^{-\phi_{B0}} 
(m - \chi_{B0} n)
H'^{-2 } \,d H \wedge d x^2 \wedge d x^3,\nonumber\\
H_5 &= & n \,\frac{\sqrt{2} \kappa_0 Q_0^3 }{\Omega_5} \,(\ast \epsilon_5 +
\epsilon_5).
\label{eq:iibthreerf}
\end{eqnarray}
In the above, the  Harmonic function $H'$ continues to be given by 
Eq.\ (\ref{eq:nhf2}) but the  Harmonic function $H$ is now
\begin{equation}
H = 1 + \frac{Q_3}{r^4},
\label{eq:hfthree}
\end{equation}
where $Q_3 = \Delta_{(m,n)}^{1/2} \sqrt{2} \kappa_0 Q_0^3 
/(4 \Omega_5)$. All the other
quantities have already been defined.

{\bf (F, D4) Bound State}\footnote{The  classical configuration 
of this bound state 
was also given in  \cite{grelpt}, obtained by  dimensional reductions from 
the (M2, M5) bound state in D = 11.}:
Repeating the same procedure the various field 
configuration for this solution are:
the metric,
\begin{eqnarray}
d s^2_{10}~ ({\rm  type~ IIA}) =&& e^{3 \phi_{B0}/8}\, H'^{3/8} \, H^{1/4}\, 
[ H^{-1}\, (- (d x^0)^2 +
(d x^1)^2) \nonumber\\
&& +  e^{-\phi_{B0}}\, H'^{-1} \,( (d x^2)^2 + (d x^3)^2 + (d x^4)^2)
+ d y^i d y^i],
\label{eq:iiamfour}
\end{eqnarray}
with $i = 1, 2, \cdots, 5$; the type IIA dilaton,
\begin{equation}
e^{\phi_A} = e^{\phi_{B0}/4}\, H'^{1/4} \,H^{- 1/2},
\label{eq:iiadfour}
\end{equation} 
and the remaining non-vanishing fields,
\begin{eqnarray}
F_3 &= &- \Delta_{(m,n)}^{-1/2} \, e^{\phi_{B0}}\, 
(m - \chi_{B0} n)\, d H^{-1}
\wedge d x^0 \wedge d x^1,\nonumber\\
F_4' &=& n \frac{\sqrt{2} \kappa_0 Q_0^4}{\Omega_4} \,\epsilon_4 + 
n \, \Delta_{(m,n)}^{-1} \,e^{-\phi_{B0}}\, (m 
- \chi_{B0} n) \,H'^{-2}\, d H \wedge d x^2 \wedge d x^3 \wedge d x^4.
\label{eq:iiarffour}
\end{eqnarray}
In the above, the Harmonic function $H'$ is as given by Eq.\ (\ref{eq:nhf2}) 
but the  Harmonic
function $H$ is now
\begin{equation}
H = 1 + \frac{Q_4}{r^3},
\label{eq:iiahffour}
\end{equation}
where $Q_4 = \Delta_{(m,n)}^{1/2} \sqrt{2} \kappa_0 Q_0^4 /(3 \Omega_4)$.

{\bf (F, D5) Bound State}: The field configurations for this bound state are: 
the Einstein metric,
\begin{eqnarray}
d s^2_{10} ~({\rm type~ IIB}) =&& e^{\phi_{B0}/2}\, H'^{1/2} \, H^{1/4} \,
[ H^{-1} (- (d x^0)^2 +
(d x^1)^2) \nonumber\\
&&+  e^{-\phi_{B0}}\, H'^{-1} ( (d x^2)^2 + (d x^3)^2 
+ (d x^4)^2 + (d x^5)^2)
+ d y^i d y^i],
\label{eq:iibmfive}
\end{eqnarray}
with $i = 1, 2, 3, 4$; the type IIB dilaton,
\begin{equation}
e^{\phi_B} = H^{- 1/2},
\label{eq:iibdfive}
\end{equation}
and the remaining non-vanishing fields,
\begin{eqnarray}
H_3 ~({\rm NSNS}) &=&  - \Delta_{(m,n)}^{-1/2}\, e^{\phi_{B0}}\, 
(m - \chi_{B0} n)\, d H^{-1}\wedge d x^0 
\wedge d x^1, \nonumber\\     
H_3 ~({\rm RR}) & = & n \,\frac{\sqrt{2} \kappa_0 Q_0^5}{\Omega_3}\, 
\epsilon_3,\nonumber\\
H_5 &=&  n  \,\Delta_{(m,n)}^{-1} \,e^{-\phi_{B0}} \,(m 
- \chi_{B0} n)\, H'^{-2} \,d H \wedge d x^2 \wedge d x^3 \wedge d x^4 
\wedge d x^5.
\label{eq:iibrffive}
\end{eqnarray}
Again the Harmonic function $H'$ remains the same as given in  
Eq.\ (\ref{eq:nhf2}) but the Harmonic
function $H$ is now
\begin{equation}
H = 1 + \frac{Q_5}{r^2},
\label{eq:iibhffive}
\end{equation}
where $Q_5 = \Delta_{(m,n)}^{1/2} \sqrt{2} \kappa_0 Q_0^5 /(2 \Omega_3)$.

{\bf (F, D6) Bound State}\footnote{The classical configuration of this bound state
was also given in \cite{cosp}.}: In this case we have the Einstein metric,
\begin{eqnarray}
d s^2_{10} ~({\rm type ~IIA}) = &&e^{5 \phi_{B0}/8}\, H'^{5/8} \,H^{1/4} \, 
[ H^{-1} (- (d x^0)^2 +
(d x^1)^2) \nonumber\\
&&+  e^{-\phi_{B0}} H'^{-1} ( (d x^2)^2 + \cdots + 
(d x^6)^2) + d y^i d y^i],
\label{eq:iiamsix}
\end{eqnarray}
with $i = 1, 2, 3$; the type IIA dilaton,
\begin{equation}
e^{\phi_A} = e^{ - \phi_{B0}/4} \,H'^{- 1/4}\, H^{- 1/2},
\label{eq:iiadsix}
\end{equation}
and  the remaining non-vanishing fields,
\begin{eqnarray}
F_2 &=& n \,\frac{\sqrt{2} \kappa_0 Q_0^6}{\Omega_2}\, \epsilon_2,\nonumber\\
F_3 &=&- \Delta_{(m,n)}^{-1/2} \,e^{\phi_{B0}}\, (m - \chi_{B0} n) 
\,d H^{-1}\wedge d x^0 
\wedge d x^1, \nonumber\\
F_4' &=& - n \,\Delta_{(m,n)}^{- 1/2} \,e^{\phi_{B0}}\, 
(m - \chi_{B0} n)\, H^{-1}\, 
\frac{\sqrt{2} \kappa_0 Q_0^6}{\Omega_2}\, d x^0 \wedge d x^1 
\wedge \epsilon_2,
\label{eq:iiarfsix}
\end{eqnarray}
Once again the Harmonic function $H'$ continues to be given by  
Eq.\ (\ref{eq:nhf2}) but the  Harmonic
function $H$ is 
\begin{equation}
H = 1 + \frac{Q_6}{r},
\label{eq:iiahfsix}
\end{equation}
where $Q_6 = \Delta_{(m,n)}^{1/2} \sqrt{2} \kappa_0 Q_0^6 /\Omega_2$.

{\bf (F, D7) Bound State}: The various field configurations 
in this case are described by the following Einstein metric,
\begin{eqnarray}
d s^2_{10}~ ({\rm type~ IIB}) =&& e^{3 \phi_{B0}/4}\,  H'^{3/4}\, H^{1/4} \,
[ H^{-1} (- (d x^0)^2 +
(d x^1)^2)\nonumber\\
&& +  e^{-\phi_{B0}} \, H'^{-1} ( (d x^2)^2 + (d x^3)^2  + \cdots + 
(d x^7)^2) + d y^i d y^i],
\label{eq:iibmseven}
\end{eqnarray}
with $i = 1, 2$; the type IIB dilaton,
\begin{equation}
e^{\phi_B} = e^{- \phi_{B0}/2}\, H'^{ -1/2} \,H^{- 1/2},
\label{eq:iibdseven}
\end{equation}
and the remaining non-vanishing fields,
\begin{eqnarray}
d \chi &=& n \,\frac{\sqrt{2} \kappa_0 Q_0^7}{\Omega_1}\, 
\epsilon_1,\nonumber\\
H_3 ~({\rm RR}) &=&  - n \,\Delta_{(m,n)}^{- 1/2}\, e^{\phi_{B0}}\, 
(m - \chi_{B0} n)\, H^{-1}\,
 \frac{\sqrt{2} \kappa_0
Q_0^7}{\Omega_1} \,d x^0 \wedge d x^1 \wedge \epsilon_1, \nonumber\\
H_3 ~({\rm NSNS}) & = & - \Delta_{(m,n)}^{-1/2}\, e^{\phi_{B0}}\, 
(m - \chi_{B0} n)\, d H^{-1}\wedge d x^0 
\wedge d x^1.
\label{eq:iibrfseven}
\end{eqnarray}
Here the Harmonic function $H'$ is again given  by  Eq.\ (\ref{eq:nhf2}) 
but the Harmonic
function $H$ is now
\begin{equation}
H = 1 - Q_7 ~{\rm ln} r,
\label{eq:iibhfseven}
\end{equation}
where $Q_7 = \Delta_{(m,n)}^{1/2}\sqrt{2} \kappa_0 Q_0^7 /\Omega_1$.

A detail discussion of the meaning and the other properties of the 
above solutions will be given in the following section. We, however,
continue this section to briefly indicate how to construct (W, Dp)
bound states in type II theories from the (F, Dp) bound states already 
constructed.

So far, we have performed T-duality only along the transverse directions
of the F-strings to obtain (F, Dp) bound states from the known (F, D1) bound
state. But we note that following the prescription given in the previous 
section, we can also T-dualize the above (F, Dp) solutions for $1\le p \le
7$ along $x^1$-direction of the F-strings.
This will give (W, Dp) bound state solutions for $0\le p \le 6$. We here 
present 
the Einstein metric, the dilaton and the Kaluza-Klein vector potential 
for each of these 
solutions. The remaining non-vanishing 
fields can be obtained easily from the corresponding ones of 
(F, Dp). For the (W, Dp) solutions we have the following Einstein metric, 
\begin{eqnarray}
d s^2_{10} =&& e^{(p + 1) \phi_{B0}/8} \,H'^{( p + 1)/8}\, 
[ - H^{ -1} (d x^0)^2 
+ e^{-\phi_{B0}}\, H'^{-1} \, H (d x^1 + {\cal A}_0 d x^0)^2 \nonumber\\
&&+ e^{-\phi_{B0}}\, H'^{ -1} (
(d x^2)^2 + \cdots + (d x^p)^2) + d y^i dy^i],
\label{eq:gwdpm}
\end{eqnarray}
with $i = 1, 2, \cdots, (9 - p)$; the dilaton,
\begin{equation}
e^\phi = e^{(3 - p)\phi_{B0}/4} \,H'^{(3 - p)/4},
\label{eq:gwdpd}
\end{equation}
and the Kaluza-Klein vector potential
\begin{equation}
{\cal A}_0 = - \Delta_{(m,n)}^{- 1/2}\, e^{\phi_{B0}} 
(m - \chi_{B0} n)\, H^{ - 1}.
\label{eq:gwdpv}
\end{equation}
In the above, the Harmonic function $H$ is given by,
\begin{equation}
H = \left\{\begin{array}{ll}
            1 + \frac{Q_{p + 1}}{r^{6 - p}},& 0 \le p \le 5,\\
            1 - Q_7~ \mbox{ln}~r, & p = 6, \end{array}
	    \right.
\end{equation}
and the Harmonic function $H'$ continues to be given by 
Eq.\ (\ref{eq:nhf2}). All the other
quantities are already given before. Note that for $m = 0$ 
(also $\chi_{B0} = 0$ and for simplicity we 
also set $\phi_{B0} = 0$\footnote{We will discuss this in the 
following section.}), 
we have $H' = H$ 
and ${\cal A}_0 = 0$.  Then the above solution represents 
an infinite periodic array of
Dp-branes along the $x^1$-axis. While for $n = 0$, $H' = 1$, and  the 
above solution 
represents a gravitational wave propagating in $x^1$-direction with 
isometries along   
$(x^1, x^2, \cdots, x^p)$-directions. We will not discuss the (W, Dp) 
bound states any
further in this paper and  focus on (F, Dp) ones for the rest.

\section{Properties of (F, D$_p$) Bound States\protect\\}
\label{sec:pfdp}

	The worldvolume picture for a (F, Dp) bound state described in 
\cite{lurone} consists of
constant electric field lines flowing along, say,  $x^1$-axis in the 
Dp-brane
worldvolume. This picture applies for each of our spacetime configurations 
of (F, Dp) given 
in the previous section. In order to see that, one can examine that for 
$m = 0$ (with
$\chi_{B0} = 0$ and $\phi_{B0} = 0$ for simplicity), the metric describes  
Dp-branes lying along ($x^1, x^2, \cdots, x^p$)-directions
while  for $n = 0$ (hence $H' = 1$), it describes F-strings lying along 
the $x^1$-direction
with isometries along $(x^1,x^2, \cdots, x^p$)-directions. So, 
this configuration indeed describes F-strings within the D-branes 
in consistency with our
worldvolume picture. In order to make such an identification more precise, 
we need to 
calculate the charges carried by the Dp-brane and by the F-strings, and also 
the mass per unit $p$-brane worldvolume, and  compare them with what we 
obtained in 
\cite{lurone} based on the worldvolume approach.

    Let us first calculate the RR charge for the Dp-brane in a (F, Dp) 
bound state
 configuration given in the previous section.
 This can be done easily using either Eq.(3.9) or Eq.(3.10) depending
on whether the charge is electric-like or magnetic-like with the 
corresponding explicit $(p + 2)$-form field strength
 given in the previous section. We   
 have $e_p = n\, Q_0^p$ for an electric-like RR charge or  
$g_p = n\, Q_0^p$ for a 
 magnetic-like RR charge. The integer $n$ originates from that of the 
D-string in the 
 (F, D1) bound state.  So the RR charge is automatically 
 quantized given the quantization of D-string charge, a well-known 
fact that the
 charge  quantization for one extended object will imply charge 
quantizations for the rest
 of the extended objects in string/M theory.  This 
 is consistent with our worldvolume result. Note that 
$Q_0^p \,Q_0^{6 - p} = 2 \pi$.
 
 As hinted earlier,  Eq.\ (\ref{eq:ecd}) cannot be applied simply to calculate 
 the electric-like charge associated with the F-strings in (F, Dp). 
The reason is, as
 explained in \cite{lurone}, that the notation `F' in (F, Dp) means actually
an infinite
 number of parallel strings in the bound state.
As mentioned in \cite{lurone}, there is one NS-string 
(or equivalently $m$ F-strings)  per $(2\pi)^{p -1}
\alpha'^{(p - 1)/2}$  of ($p - 1$)-dimensional area. We will show later in 
this section that this is
indeed true.
 In 1 + 3 
 dimensional electrostatics, we know that in order to use Gauss law to obtain
 the charge per unit length for a uniform line distribution of charge, 
we have to choose
 a cylinder (with the line charge at the center) as the Gauss surface 
rather than a 2-sphere as for a
 point charge. For the present case, we therefore should choose the 
integration in 
 Eq.\ (\ref{eq:ecd}) over $R^{p -1} \times S^{8 - p}$ rather than over 
an asymptotic
 $S^7$. We can therefore have charge per unit $(p -1)$-dimensional area 
for the F-strings.
 
	In order to give meaningful calculations for this quantity and for the 
mass per unit Dp-brane worldvolume which will be used to determine the tension
 for a (F, Dp) bound 
state, we need a good asymptotic behavior for the metric such that the 
$(p - 1)$-dimensional area and the worldvolume can be defined with 
respect to certain frame
metric. This is also necessary for us to make comparisons with the results 
obtained in \cite{lurone} based on the worldvolume analysis. 
Let us take a close look at each
Einstein frame 
metric for those (F, Dp) bound state configurations. They, 
along with the corresponding 
dilatons, can
actually be expressed in a unified way as
\begin{eqnarray}
 d s^2 = &&e^{(p - 1) \phi_{B0}/8}\, H'^{(p - 1)/8}\, H^{1/4}\,
[ H^{ -1} (- (d x^0)^2 +
 (d x^1)^2) \nonumber\\
 &&+ e^{- \phi_{B0}}\, H'^{-1} \,( (d x^2)^2 + \cdots 
+ (d x^p)^2) + d y^i d y^i],
 \label{eq:gem}
 \end{eqnarray} 
 for the metric with $i = 1, 2, \cdots, 9 - p$;  and
\begin{equation}
e^\phi = e^{(5 - p)\phi_{B0}/4} \,H'^{(5 - p)/4} \, H^{ - 1/2}.
\label{eq:gd}
\end{equation}
 for the dilaton. In the above, the Harmonic function $H$ is 
\begin{equation}
H = 1 + \frac{Q_p}{r^{7 - p}},
\label{eq:ghf}
\end{equation}
with 
\begin{equation}
Q_p = \frac{\Delta_{(m, n)}^{1/2} \sqrt{2} \kappa_0 Q_0^p} 
{(7 - p) \Omega_{8 - p}},
\label{eq:gcc}
\end{equation}
and the Harmonic function $H'$ is given by Eq.\ (\ref{eq:nhf2}).

 Note first that  the constants
 $\phi_{B0}$ and $\chi_{B0}$ appearing in all the above solutions 
are no longer the 
 asymptotic values for the dilaton and the axion except for the original 
(F, D1) bound state
 configuration. Actually, the axion for $p = 3, 5, 7$ has nothing 
to do with the constant
 $\chi_{B0}$. In fact, $\chi_{B0}$ is the asymptotic value of the 
$(p - 1)$-form RR gauge 
 potential 
 in the corresponding (F, Dp) configuration. The asymptotic value of 
dilaton for every $p$ 
 except for $p = 5$ 
 is related to the original $\phi_{B0}$ but not equal. 
This can be understood from the fact
 that a T-duality transformation shifts the dilaton value\cite{bus}. From 
 the dilaton expression for each of the above solutions, we can see 
that the asymptotic 
 value of the dilaton is reduced by a quarter of $\phi_{B0}$ for 
every T-duality 
 transformation. For $p = 5$, the asymptotic value vanishes.  Second, 
even though 
 we start with an Einstein metric for (F, D1) which is 
asymptotically Minkowski, none of
 the metrics derived by T-duality transformations remains to be so. 
Even worse, none of them
 remains asymptotically  as a constant scaling factor times Minskowski metric. 
However, 
 there exists a constraint on the asymptotic behavior of the 
Einstein metric for the
 (F, D1) bound state configuration which preserves not only the 
asymptotic behavior for each
 of the derived Einstein metrics but also the asymptotic value for the 
dilatons\footnote{This
 remains true also for the (W, Dp) bound states.}. This turns
 out to be the one which imposes the corresponding string metric 
rather than the Einstein
 metric to be asymptotically Minkowski. This constraint has been 
used in the literature and 
 once again we see its significance. We will discuss possible 
reasons behind  this and other
 related issues in a separate paper. 
 
 Once we make such a choice for the asymptotic metric, 
Eq.\ (\ref{eq:gem}) and (\ref{eq:gd})
 become
 \begin{eqnarray}
 d s^2 = && e^{- \phi_{B0}/2} \,H'^{(p - 1)/8} \,H^{1/4}\,\left[ H^{ -1} 
(- (d x^0)^2 + (dx^1)^2)\right. \nonumber\\
 && \left. + H'^{-1} ( (d x^2)^2 + \cdots + (d x^p)^2) + d y^i d y^i
 \right],
 \label{eq:ngem}
 \end{eqnarray} 
 and
\begin{equation}
e^\phi = e^{\phi_{B0}}\, H'^{(5 - p)/4}\, H^{ - 1/2}.
\label{eq:ngd}
\end{equation}
The Harmonic function $H$ is given by Eq.\ (\ref{eq:ghf}) but with
 $Q_p \rightarrow e^{3 \phi_{B0}/2} Q_p$ with 
$Q_p$ given in Eq.\ (\ref{eq:gcc}).
 The expressions (4.5) and (4.6) have the above mentioned properties. 
For example, the asymptotic value for the
 dilaton  is always $\phi_{B0}$, the one in the original (F, D1) bound state. 
In what
 follows, we simply use $\phi_0$ rather than $\phi_{B0}$ as the 
asymptotic value for the
 dilaton. 
  We give below the 
explicit  form of NSNS 3-from field strength\footnote{$G_3 = F_3$ 
in type IIA while 
$G_3 = H_3 ~({\rm NSNS})$ in type IIB.} $G_3$ for the purpose of 
calculating the 
charge per unit $(p-1)$-dimensional area for F-strings  which is
\begin{equation} 
G_3 = - \Delta_{(m, n)}^{- 1/2} \,e^{\phi_{0}/2} \,(m - \chi_{B0} n) \,
d H^{ -1}\wedge d x^0
\wedge d x^1.
\label{eq:gthreef}
\end{equation}

As discussed above, the charge per unit $(p-1)$-dimensional area for the
F-strings should be calculated as,
\begin{equation}
e_1 = \frac{1}{\sqrt{2} \kappa_0} \int_{R^{p - 1}\times S^{8 - p}} 
(e ^{ - \phi} \ast G_3
+ \cdots),
\label{eq:gcfs}
\end{equation}
where $\cdots$ indicates possible non-vanishing Chern-Simons terms. 
Actually, the 
Chern-Simons terms do contribute to this charge.
The charge $e_1$ itself must be infinite  since we have an infinite 
number of F-strings.
But we expect that $e_1 / A_{p -1}$ should be finite with 
$A_{p - 1} = \int d x^2\wedge
\cdots \wedge d x^p$ the coordinate $(p -1)$-dimensional area. Indeed, 
we find
\begin{equation}
|e_1|/A_{p -1} = m  Q_0^p,
\label{eq:gcpfs}
\end{equation}
As mentioned in \cite{lurone}, we have one NSNS-string 
(or $m$ F-strings) per
$(2\pi)^{p -1} \alpha'^{(p -1)/2}$ of ($p -1$)-dimensional area  
from the worldvolume 
point of view. So, in order to compare with our interpretation for the
F-strings along $x^1$-direction  we must multiply  the left side 
of the above equation
with this $(p -1)$-dimensional area. Thus we obtain,
\begin{equation}
(2 \pi)^{p -1} \alpha'^{(p -1)/2} \frac{|e_1|}{ A_{p -1}}
  =  \sqrt{2} \kappa_0 \,m  T_f,
\label{eq:conf}
\end{equation}
Now in order for the left hand side of Eq.(4.10) to represent the
charge of F-strings per $(2\pi)^{p-1} \alpha'^{(p-1)/2}$ of 
$(p-1)$-dimensional area, we must replace the coordinate area $A_{p-1}$
by the one measured in string metric asymptotically. But, since we
have chosen our string metric to be asymptotically flat, they are the same.
Note also that according to our definition the charge $e_p$ of the
$p$-brane is related to the corresponding tension $T_p$ as $e_p =
{\sqrt 2} \kappa_0 T_p$ and therefore, Eq.(4.10) clearly states that
there are indeed $m$ F strings in the (F, Dp) bound states per $(2\pi)^{p-1}
\alpha'^{(p-1)/2}$ of $(p-1)$-dimensional area, confirming our interpretation.

There is another way to obtain the above result which gives a direct link 
between the 
spacetime F-strings and the worldvolume electric flux lines. 
This will be important for 
our discussion of the $AdS_5/CFT_4$ correspondence given in section 5. 
The low-energy
effective field theory on a Dp brane worldvolume contains a coupling,
\begin{equation}
\frac{1}{T^p_0 g} \int d^{p + 1} \sigma B_{\mu\nu} (2\pi \alpha' F^{\mu\nu}),
\label{eq:coupling}
\end{equation}
where $B_2$ is the pull-back of the NSNS 2-form potential on the worldvolume, 
$F$ is the worldvolume gauge field strength\footnote{Strictly
speaking, we should use ${\cal F} = F - B_2$ instead.}, 
$T_0^p = 1/[(2\pi)^p \alpha'^{(p + 1)/2}]$ is the $p$-brane tension units 
and $g = e^{\phi_0}$ is
the string coupling. Because of this term the equation of motion 
takes the form (in terms of our
conventions)\footnote{Here, 
we do not include an F-string source to the equation
of motion since we are considering an infinite number 
of parallel F-strings in the bulk.}
\begin{equation}
d (e^{-  \phi}\ast G_3 + \cdots ) = \frac{ 2 \,\kappa_0^2}
{ g \,T_f \,T^p_0} \ast F \wedge \delta^{9 - p},
\label{eq:ceom}
\end{equation}
where $\ast$ in $\ast G_3$ denotes the Hodge dual in $D = 10$ spacetime while 
the $\ast$ in $\ast F$ denotes the Hodge dual within the 
$(p + 1)$-dimensional worldvolume.
In the above, $\delta^{9 - p}$ denotes a $(9 - p)$-form delta function 
on the Dp brane
worldvolume. We use this equation for the purpose of relating 
the F-string charge to the 
 electric flux of gauge field on the worldvolume.  
 (Note that a similar equation has been used in \cite{str} for the
charge conservation.) In order to get the charge of F-strings in (F, Dp)
bound state, we should
 integrate the above equation on $R^{p - 1} \times R^{9 - p}$.  We have
 \begin{eqnarray}
 \int_{R^{p - 1} \times R^{9 - p}} d (e^{- \phi} \ast G_3 + \cdots) &=& 
 \int_{R^{p - 1} \times S^{8 - p}} (e^{-\phi} \ast  G_3 + \cdots ),\nonumber\\
 &=& \frac{ 2\, \kappa_0^2}{ g \,T_f \,T^p_0}  \int_{R^{ p - 1}} \ast F.
 \label{eq: iceom}
 \end{eqnarray} 
This gives, 
 \begin{eqnarray}
 e_1 &=& \frac{1}{\sqrt{2} \kappa_0} \int_{R^{p - 1}\times S^{8 - p}} 
 (e ^{ - \phi} \ast G_3 + \cdots), \nonumber\\
 &=& \frac{ \sqrt{2} \kappa_0}{ g T_f T^p_0}   F_{01} A_{p - 1},
 \label{eq:cfr}
 \end{eqnarray}
 where we have taken $F_{01}$ as the only non-vanishing constant component
of the $U(1)$ gauge field strength on the worldvolume of Dp-brane. 
If we denote $Q_1$ as the
 charge of F-strings per $(2\pi)^{p -1} \alpha'^{(p -1)/2}$  area, we
 have
 \begin{equation}
 \frac{Q_1}{\sqrt{2}\, \kappa_0} = F_{01} / g,
 \label{eq:cfrone}
 \end{equation}
 where $Q_1/\sqrt{2} \kappa_0$ is the total tension of these F-strings. 
So we establish the precise
 relationship between the F-strings and the electric flux lines. If we take 
 $Q_1/\sqrt{2} \kappa_0 = m T_f$ from the above calculated value, we 
have $F_{01} = g m T_f$ which
 agrees precisely with what we obtained in \cite{lurone}. Certainly, the other
 way around is also true.

We now use the metric Eq.\ (\ref{eq:ngem}) to calculate the mass per unit 
$p$-brane volume
for a (F, Dp) bound state and compare the result with what we obtained 
in \cite{lurone}.
In doing so, we need to generalize the ADM formula 
in [18] to accommodate the following
metric,
\begin{equation}
d s^2 = - A(r) dt^2 + B(r) d r^2 + r^2 C (r) d \Omega_{D - p - 2} + 
D (r) \delta_{ij} d x^i d x^j + E (r) \delta_{kl} d x^k d x^l,
\label{eq:gm}
\end{equation}
where $D$ is the spacetime dimensions, $p$ is the spatial dimensions of a 
$p$-brane, and
indices $(i, j = 1, 2, \cdots, m)$ and $(k, l = m + 1, \cdots, p)$. If 
$ A(\infty) = B(\infty) = C(\infty) = D (\infty) = E (\infty) = a_p$ 
(with $a_p$ a constant)
[19], we have the ADM mass per unit $p$-brane volume as
\begin{eqnarray}
M_p = - \frac{\Omega_{\tilde{d} + 1}}{2 \kappa^2_0} [ &&
(\tilde{d} + 1) \,r^{\tilde{d} + 1}\partial_r C(r) + m \,r^{\tilde{d} 
+ 1} \partial_r D (r)
+ (p - m) \,r^{\tilde{d} + 1} \partial_r E (r) \nonumber\\
&&- (\tilde{d} + 1)\, r^{\tilde{d}} 
\left(B (r) - C (r)\right) ]_{r \rightarrow \infty},
\label{eq:gadm}
\end{eqnarray}
with $\tilde{d} = D - p - 3$. 
Applying  the above formula to our case, we get the mass per unit 
$p$-brane volume for 
a (F, Dp) bound state as
\begin{equation}
M_p = g \sqrt{n^2 + g^2 (m - \chi_{B0} n)^2} ~T^0_p,
\label{eq:gmass}
\end{equation}
where we have used 
$Q_p = \Delta_{(m,n)}^{1/2} e^{3\phi_{0}/2} \sqrt{2}\kappa_0 Q_0^p/[(7 - p) 
\Omega_{8 - p}]$, 
the expressions for $Q_0^p$ from Eq.\ (\ref{eq:puc}) and 
the $\Delta$-factor from 
Eq.\ (\ref{eq:df}), respectively. Here $T^p_0 = 1/[(2\pi)^p 
\alpha'^{(p + 1)/2}]$ is the $p$-brane
tension units and $g = e^{\phi_{0}}$ is the string coupling. We can relate the 
$M_p$ and the tension $T_p (m, n)$ of (F, Dp) in many ways. 
{}For example, they can be related by looking at the scaling behavior of the 
energy-momentum
tensor due to the contribution of the Dp-brane worldvolume action 
under a constant rescaling
of the spacetime metric. But we here use a very simple approach. 
We know that $M_p$ must be
proportional to $T_p (m, n)$ and the proportionality constant must be 
some power of the constant 
$a_p$, i.e., $M_p = a_p^\alpha T_p (m, n)$ with $\alpha$ 
an as yet undetermined constant.
Note that here $a_p = g^{- 1/2}$. When $m = 0, \chi_{B0} = 0$,
we know $T_p (0, n) = n T^p_0/ g$ and from Eq.\ (\ref{eq:gadm}) 
we also have 
$M_p = g n T^p_0$. 
So we
have  $\alpha = - 4$ and we get $M_p = g^2 T_p (m, n) $ in general. 
Using this relation and Eq.(4.14) we find
the tension $T_p (m, n)$ for the (F, Dp) bound state as
\begin{equation}
T_p (m, n) = \frac{1}{g} \sqrt{n^2 + (m - \chi_{B0} n)^2 g^2}~ T^p_0,
\label{eq:gtension}
\end{equation}
which agrees precisely with what we obtained in \cite{lurone} for $\chi_{B0} = 0$. This also
shows that (F, Dp) is a non-threshold bound state.

\section{Decoupling Limit for the (F, D3) bound state\protect\\}
\label{sec:ads/cft}

	The properties of the (F, Dp) bound states studied in the 
previous section are either the 
BPS ones  or the consequences of 
BPS properties. Therefore, it is not surprising that we find agreements 
in both the spacetime and 
the worldvolume approaches even though, strictly speaking, 
a flat background is 
always assumed in the worldvolume study for these bound states 
in \cite{lurone}. 
These calculations clearly demonstrate that our spacetime 
bound state configurations (F, Dp) are identical to those bound states 
obtained from the 
worldvolume study in\cite{lurone}. The F-strings in (F, Dp) should be 
identified with the 
electric flux or field lines in the corresponding gauge theory 
living on the worldvolume. 
Now the question is: Can the same $AdS/CFT$ correspondences 
conjectured by Maldacena
in \cite{mal} for $p = 3$ and by Itzhaki et al in \cite{itzmsy} 
for $p \neq 3$
be proposed in a 
similar fashion based on the (F, Dp) bound states rather than  
on the simple Dp branes? If so,
can we learn anything new? We will examine these for the $p = 3$ case
in the following. 
The analysis for $p \neq 3$ 
can  be made similarly as for the $p = 3$ case along the
line given in \cite{itzmsy}.

In general, when the so-called D-branes appear in string theory, 
we need to consider not only
the modes that propagate in the bulk but also the modes that 
propagate on the D-branes. The modes
propagating on the D-branes are associated with open strings ending 
on the D-branes. In general,
these modes interact not only among themselves but also with the 
modes propagating in the bulk.
However, there exists a limit that decouples the
modes propagating on the branes from the modes propagating in the 
bulk and is also typically 
a low energy limit. The latter says that the modes on the brane are 
just the massless ones of 
the open strings. In this limit, the D-brane theory  becomes the 
corresponding SYM theory (for 
$p \le 3$). 

In the case of $p = 3$, the D3 brane theory is described by 
${\cal N} = 4$ SYM
in $1 + 3$ dimensions under this limit. But  the D3 brane configuration is 
 also described in terms of the metric
and other fields in the bulk. The so-called decoupling or field 
theory limit is
\begin{equation}
g_{\rm YM}^2 = 2\pi g = {\rm fixed},~\qquad \alpha' \rightarrow 0,
\label{eq:dcl}
\end{equation}
where $g = e^{\phi_0}$ is the string coupling constant and 
$g_{\rm YM}$ is the Yang-Mills coupling constant. Note that
the ten dimensional Newton constant $\kappa^2 \sim g^2 \alpha'^4$ 
vanishes in this limit as it should be.
Under this limit,  the BPS D3-brane configuration in the bulk becomes  
flat Minkowski for any fixed non-zero
 $r$. Therefore,  to have something
nontrivial, we need to consider an additional limit,
\begin{equation}
U \equiv \frac{r}{\alpha'} = {\rm fixed},~\qquad \alpha' \rightarrow 0.
\label{eq:esl}
\end{equation}
The above says that we keep the mass of the stretched strings 
between D3 branes fixed. This $U$ actually
sets the energy scale in the field theory since it is  the expectation 
value of the Higgs. This also
implies that we are considering finite  energy configurations 
in the field theory. As discussed in
\cite{mal}, if we consider $n$ parallel  D3 branes, then the above 
decoupling limits say that we 
are bringing the branes together but the Higgs expectation value 
corresponding to this separation
remains fixed. The resulting theory on the brane is ${\cal N} = 4$ $U(n)$ 
SYM theory in $1 + 3$
dimensions. In what follows, we simply call 
Eqs.\ (\ref{eq:dcl}) and \ (\ref{eq:esl}) as the decoupling
limit.

	Under the  decoupling limit, the string metric describing 
the D3 branes becomes
$AdS_5 \times S^5$. The isometry of $AdS_5$ is $SO(2, 4)$ 
and this  is also the conformal group
in $1 + 3$ dimensions. We also have isometries of $S^5$ as 
$SO(6)\sim SU(4)$. This symmetry is
identical to the R-symmetry of the ${\cal N} = 4$ SYM. After 
including fermionic generators 
required by supersymmetry, we have the full isometry supergroup 
$SU(2, 2\mid 4)$ for the 
$AdS_5 \times S^5$ background, which is identical to the 
${\cal N} = 4$ superconformal group.
We would like to emphasize that we have the conformal symmetry 
$SO(2,4)$ only at $U = 0$. For any fixed
non-zero $U$, this symmetry is spontaneously broken since 
$U$ represents the expectation value of the
Higgs.

To validate the configuration of D3 branes as a stringy one, 
the large $n$ limit has to be taken given
the fixed but small string coupling\footnote{For large $g$, 
the D-string frame is chosen.}.
It is well-known that there is an SL(2,Z) strong-weak duality symmetry 
in both type IIB supergravity and 
${\cal N} = 4$ SYM. These symmetry identifications, among other things, 
led Maldacena to  conjecture
that the large $n$, ${\cal N } = 4$ SYM theory with gauge group 
$U(n)$ is actually equivalent to the
ten dimensional type IIB supergravity on $AdS_5 \times S^5$. 
However, the supergravity itself is not
a consistent quantum theory. On the other hand, the SYM is a 
unitary quantum theory. Moreover, the above 
symmetry identifications are independent of the large $n$ limit 
(even though it is needed to validate the 
solution). These two facts 
led Maldacena further to conjecture that type IIB string on 
$AdS_5 \times S^5$ is equivalent to
the ${\cal N} = 4$ SYM theory  for a general $n$.

Let us now apply the same process to the (F, D3) configuration 
obtained in section 3, i.e., 
taking the decoupling limit given by Eqs.\ (\ref{eq:dcl}) and \ (\ref{eq:esl}). In other words, 
we should 
examine the near-horizon geometry of (F, D3) configuration with the 
string coupling $g$ fixed\footnote{For simplicity, we set $\chi_{B0} = 0$ 
from now on.}.
The 
near horizon string-frame metric in this case is
\begin{eqnarray}
d s^2 =&& \alpha' \left[ \frac{U^2}{\sqrt{4\pi n g}} \left(\frac{ n^2
e^{-\phi_0}}{\Delta_{(m,n)}}\right)^{3/4} 
( - (d x^0)^2 + (d x^1)^2 ) \right.\nonumber\\
&& + \frac{U^2}{\sqrt{4\pi n g}} \left(
\frac{ n^2 e^{-\phi_0}}{\Delta_{(m,n)}}\right)^{ - 1/4} 
((d x^2)^2 + (d x^3)^2)
\nonumber\\ 
&& \left. + \sqrt{4 \pi n g} \left(\frac{ n^2
e^{-\phi_0}}{\Delta_{(m,n)}}\right)^{1/4} \left(\frac{d U^2}{U^2} + 
d \Omega_5^2 \right) \right],
\label{eq:nhg}
\end{eqnarray}
 and the dilaton is
\begin{equation}
e^\phi = \left(n^2 e^{-\phi_0} /\Delta_{(m, n)}\right)^{1/2} e^{\phi_0}.
\label{eq:dthree}
\end{equation}
It is clear from (5.3) and (5.4) that unlike the simple D3 brane, the  
near-horizon geometry in this case is
 not automatically $AdS_5 \times S^5$ and the effective string coupling, 
$e^\phi$,
 is still a finite constant but is less than the string 
coupling\footnote{The dilaton 
 for (F, D3) is actually not a constant in general, nevertheless
 it is bounded as 
  $ \left(n^2 e^{-\phi_0} /\Delta_{(m, n)}\right)^{1/2} 
e^{\phi_0}< e^\phi < e^{\phi_0}$, 
where the upper bound is obtained at $r = \infty$ and the 
lower bound  at $r = 0$.} $g$. 

The above features may be expected given the appearance of an infinite 
number of F-strings in
(F, D3) and the non-threshold bound state nature of this configuration. 
On the other hand,
 in the decoupling
 limit, we expect that D3 brane  can be described equivalently 
either by the SYM theory
 on the brane or by the type IIB supergravity (or string) 
in the bulk. Depending on whether we consider 
excitations with respect to the simple D3-brane vacuum or the 
(F, D3) bound state (chosen as a new vacuum),
we have two descriptions which are different in appearance but 
probably equivalent in essence. Let us first
discuss the possible underlying IIB string/SYM correspondence 
for each of the above descriptions in 
the strong sense, i.e., not taking the large $n$ limit. We will discuss 
the large $n$ limit in the end.

If we choose the simple  BPS D3 brane configuration in the bulk 
or equivalently no excitations in the SYM 
on the brane as the vacuum, then the F-strings in the bulk or 
equivalently the electric flux lines 
in the
SYM should be formed due to excitations with respect to this vacuum. 
We know that 
the energy per unit 3-brane volume associated with the F-strings 
in the bulk or the energy density 
associated with 
the electric flux lines on the brane is finite. This implies that we 
have an infinite amount of energy
associated with either the F-strings or the electric flux lines. 
In other words, we should have $U = \infty$,
which implies that the original conformal symmetry group 
$SO(2,4)$ must be broken spontaneously to some
smaller symmetry group consisting of at most some translational 
and rotational symmetries. It is not difficult
to figure out the origin of $U = \infty$ if we recall the method we used 
in \cite{lurone}
 to find the (F, D3) bound state. As discussed in \cite{lurone}, 
the charge conservation must imply that the 
infinite parallel F-strings in (F, D3) originate from F-strings 
along the radial coordinate $r$ ending on 
 the $x^2x^3$-plane placed at $x^1 = - \infty$
 \footnote{This is chosen in order to preserve the SO(6) 
isometry of $S^5$.}. 
 These F-strings ending on $x^2x^3$ plane are the open strings, 
each semi-infinitely long, giving $U = \infty$. Inside the bulk, we
 have F-strings along $x^1$-axis, therefore we expect an $SO(1,1)$ 
symmetry. Because of the ending or the charges on 
$x^2x^3$-plane placed at $x^1 = - \infty$, we expect that 
$SO(2,4) \rightarrow SO(1, 1) \times SO(2 )$ for this 
near-horizon geometry. Thus we find that we actually have $SO(1,1) 
\times SO(2 ) \times SO(6)$ isometries for
the  near-horizon geometry of the (F, D3) bound state for the 
choice of vacuum. On the SYM side, we have
the same answer. Dilation and the special conformal transformations 
are broken spontaneously for the same reason
mentioned above. Only the subgroup $SO(1,1) \times SO(2)$ of 
$SO(2, 4)$ will leave $F_{01} = g m T_f$ invariant.
So, following the similar argument as for the $AdS_5/CFT_4$ correspondence, 
we may suggest that the type IIB string on this 
geometry is equivalent to the ${\cal N } = 4$ SYM theory in a state 
with quantized electric flux lines.

 Let us
investigate how the (F, D3) bound state emerges as our vacuum 
for the second description. 
Examining carefully the metric and the dilaton, we can see that  
$ n^2 e^{-\phi_0} /\Delta_{(m, n)}$ 
factor is the source which causes
 the near-horizon geometry to be not automatically $AdS_5 \times S^5$ 
and the effective string coupling  
 to be less than the string coupling $g$. Explicitly, we have this factor
\begin{equation} 
\frac{n^2 e^{-\phi_0}}{ \Delta_{(m,n)}} = 
\left(1 + \left(\frac{g m}{n}\right)^2
 \right)^{-1}.
\label{eq:fac}
\end{equation}
Now if we rescale the 
 coordinate time $x^0$ and the coordinate $x^1$ by a factor 
 $ \left(n^2 e^{-\phi_0} /\Delta_{(m, n)}\right)^{-1/2}$, 
then in terms of the new coordinates the 
 near-horizon geometry
 is $AdS_5 \times S^5$. So, if we rewrite the
 metric in terms of the effective string coupling 
 \begin{equation}
 g_{\rm eff} \equiv e^\phi =  \frac{g}{\left(1 + 
\left(\frac{g m}{n}\right)^2
 \right)^{1/2}},
 \label{eq:esc}
 \end{equation}
 we have
 \begin{equation}
d s^2 = \alpha' \left[ \frac{U^2}{\sqrt{4\pi n g_{\rm eff}}} 
\left( - (d x^0)^2 + (d x^1)^2 + (d x^2)^2 + (d x^3)^2 \right) +
\sqrt{4 \pi n g_{\rm eff}}  \left(\frac{d U^2}{U^2} + 
d \Omega_5^2 \right) \right].
\label{eq:enhg}
\end{equation}
 The radius for the $AdS_5$ and the $S^5$ is now given as
 \begin{equation}
 R^2/\alpha' = \sqrt{4\pi n g_{\rm eff}},
 \label{eq:radius}
 \end{equation}
  The information about the F-strings in the (F, D3) 
 bound state disappears in the above metric. Since the resulting  
near-horizon geometry differs from
 that of a simple D3 brane with 5-form flux $n$ only in the string 
coupling, we expect that the effects of 
 the F-strings are encoded in the effective string coupling $g_{\rm eff}$.
 We will see that this is indeed true.
 With respect to the new coordinates, we have $SO(2,4) \times SO(6)$ 
isometries in the bulk
 at least in appearance. How can we reconcile this with what we just 
discussed above on the SYM
 side? Moreover, what is the physics behind  such a rescaling? 
 
 In the case of simple D3 branes, the string
 coupling $g = e^{\phi_0}$ remains the same whether we are at 
$r = \infty$ or at $r = 0$. 
  This fact enables us to fix the gauge coupling
as $g_{\rm YM}^2 =  2 \pi g$ for the SYM theory describing 
the D3 branes in the decoupling limit. 
We do not expect that such a relation will be modified for 
any finite energy excitations\footnote{
 But it could be modified
if the energy involved is infinite, for example, the case we are presently
studying. If this happens, the SYM theory
may no longer be valid in describing the underlying physics in general and 
string theory should be used instead.
However, our discussion above for the first description may still 
be valid since it is based on the BPS
(F, D3) configuration.}.  However, the story here is different. 
The gauge coupling should, in the present case, be 
related to $g_{\rm eff}$ rather than 
$g$ as $g_{\rm YM}^2 = 2\pi g_{\rm eff}$. 

With respect to the metric in Eq.\ (\ref{eq:enhg}), we may say that 
a simple D3 brane with 5-form flux $n$ plays 
as an ``effective" vacuum configuration in the bulk with the 
string coupling $g_{\rm eff}$. This simple D3 brane
is also described by a ${\cal N} = 4$ $U(n)$ SYM theory in 1 + 3 
dimensions with gauge coupling 
$g_{\rm YM}^2 = 2\pi g_{\rm eff}$. This is just the usual $AdS_5/CFT_4$ 
correspondence except we have a new 
coupling constant. 
 Let us dig out the mystery behind the above ``effective" vacuum. 
We expect that the tension for the simple
 D3 brane is $n T^3_0 / g_{\rm eff}$. If we use the explicit 
expression for $g_{\rm eff}$ from 
 Eq.\ (\ref{eq:esc}), this tension is
 \begin{equation}
 \frac{n T_0^3} {g_{\rm eff}} = \frac{T_0^3}{g} \sqrt{n^2 + g^2 m^2},
 \label{eq:ident}
 \end{equation}
 which is just the tension for the (F, D3) bound state. So this 
``effective" vacuum is nothing but 
 the (F, D3) bound state. As anticipated, the effects of the F-strings 
in (F, D3)
 is encoded into the string coupling constant.

	On the SYM side, we have already hinted that the gauge coupling 
is now given by 
$g_{\rm YM}^2 = 2 \pi g_{\rm eff}$. In the linear approximation, 
\begin{equation}
\frac{1} {g_{\rm eff}} = \frac{1}{g} + \frac{1}{2 g} 
\left(\frac{ g m}{n}\right)^2,
\label{eq:lgc}
\end{equation}
where the second term for $n = 1$ is nothing but the contribution 
from the energy density of the electric
flux lines. In other words, the effect of the energy of the 
quantized electric flux lines is absorbed into the 
gauge coupling. This is in accordance with our spacetime picture 
just discussed.
 Therefore, the resulting SYM theory should
be the one describing the excitations with respect to the (F, D3) 
bound state. The superconformal symmetry 
should then be
restored. We therefore have a consistent picture both in the 
bulk and on the brane in the decoupling limit.

Our study lends further support for Maldacena's $AdS_5/CFT_4$ correspondence 
 and shows that it holds true 
even for the 
non-trivial D3 brane configurations. The new input for this correspondence 
is that we should use the effective string coupling constant 
instead of the usual one.

The two descriptions studied  may be equivalent but the picture 
for the first one is surely complicated. The
reason for this is our improper choice of the vacuum state. 
If we take the large $n$ limit, i.e.,
$n \gg m$ for fixed $g$, the effective string coupling becomes the 
usual string coupling $g$. The F-strings in the
(F, D3) now play a minor role with respect to the D3 branes which
can be seen from the tension formula
$T_3 (m, n) = n/g \left[ 1 + (m /n)^2 g^2 /2\right] T^3_0$. 
On the SYM side, on the other hand, we have the gauge coupling 
$g_{\rm YM}^2 = 2 \pi g (1 - (m/n)^2 g^2 /2)$ which implies that 
electric flux lines plays the minor role.  
 Thus the $AdS_5/CFT_4$
correspondence with respect to (F, D3) bound state is simply reduced 
to that of the 
simple D3 brane configuration. We also expect that the full 
isometry supergroup $SU(2,2\mid 4)$ in the bulk
and the superconformal symmetry in the SYM theory are restored under 
this limit in our first description above.
This description is therefore also reduced to the usual $AdS_5/CFT_4$ 
correspondence. This large $n$ limit also
validates the (F, D3) configuration as a stringy one.

Let us  now comment on the effective string coupling $g_{\rm eff}$ given by 
Eq.\ (\ref{eq:esc}) in the near-horizon region. It is clearly quantized 
in terms of relatively
prime integers $n$ and $m$ and is always less than the string 
coupling $g$. In particular, in the limit $m \gg n$, 
it becomes $g_{\rm eff} = n /m
\ll 1$, independent of the string coupling $g$. In other words, the 
effective string coupling in the 
near-horizon region is completely determined by integer $n$, the 
3-brane charge, and $m$,
the number of F-strings per $(2\pi)^2 \alpha'$ area over the 
$x^2x^3$-plane, in this limit.

We will continue this study in a more  general
D3-brane configuration, namely, ((F, D1), D3) bound state in a 
forthcoming paper \cite{lurthree}.

\section{Conclusion\protect\\}
\label{sec:c}
To summarize, we have constructed, in this paper, explicitly the (F, Dp) 
non-threshold bound state configurations for 
$2 \le p \le 7$, starting from the known (F, D1) configuration of Schwarz 
by T-duality
transformation along the transverse directions of the strings. We  have 
also presented the solutions 
for (W, Dp) non-threshold bound state configurations for 
$0 \le p \le 6$ by T-duality 
transformations on the newly obtained (F, Dp) ones along the 
longitudinal direction of F-strings 
in (F, Dp). We have shown explicitly that there are $m$ F-strings per
$(2\pi)^{p - 1} \alpha'^{(p - 1)/2}$ of $(p - 1)$-dimensional area 
which agrees with our previous result
based on the worldvolume study. We have  also calculated the tensions 
for the (F, Dp) bound states which
once again agree with our previous results. These two facts  
confirm our assertion made in our earlier paper that the bound state 
studied in \cite{lurone}
can indeed be identified with the (F, Dp) bound states here.  All of 
these bound states preserve one
half of the spacetime supersymmetries. 

	From the D-brane worldvolume point of view, each of these 
bound states consists of
a Dp-brane carrying certain units of quantized electric flux or 
field line. But from the
spacetime point of view, we have type II strings (either type IIA or 
IIB strings) lying along 
one direction in the corresponding Dp-brane worldvolume. In this paper, 
we have shown that for
each of the bound states considered,  the
type II strings can be identified with the electric flux lines 
in the gauge theory
living on the corresponding Dp brane worldvolume. In the $p = 3$ case, 
we have studied the corresponding
decoupling limit and found that the $AdS_5/CFT_4$ correspondence may 
still hold true but now with
respect to the (F, D3) bound state with an effective string coupling constant. 
The F-strings in the (F, D3) bound 
state or the electric flux lines in the corresponding SYM theory play the 
role of reducing the respective 
coupling constant in general. The string coupling in the near-horizon 
region is found to be quantized in terms of the relatively
prime integers $n$, the 3-brane charge, and $m$, the number of 
F-strings per $(2 \pi)^2 \alpha'$ area over the
$x^2x^3$-plane. It becomes independent of the asymptotic 
string coupling $g$ in the limit $m \gg n$.

\acknowledgments
We would like to thank Mike Duff for reading the manuscript and Chris Pope
for discussions. JXL acknowledges the support of NSF Grant PHY-9722090.


\begin{references}
\bibitem{ref} Since Ricci tensor is inert under an overall constant 
rescaling of the metric, it
will not 
change the total energy-momentum 
pseudotensor. Therefore, such a rescaling of metric will not change 
the mass per unit coordinate 
$p$-brane volume (In the present case, this coordinate volume 
is also the $p$-brane volume 
measured in string metric at the spatial infinity of the transverse space.).
\end{references}
\end{document}